\documentclass[aps,prb,reprint,superscriptaddress]{revtex4-2}
\usepackage[dvipdfmx]{graphicx}
\usepackage{amsmath,mathcomp}
\usepackage{physics}
\usepackage{color}
\usepackage{ulem}

\begin{document}
\title{CE-antiferromagnetic electronic structure in LaSr$_2$Mn$_2$O$_7$ revealed by micro-focused angle-resolved photoemission spectroscopy and tight-binding models}



\author{Yasutaka Sawata}
\affiliation{Department of Applied Physics, Tokyo University of Science, Katsushika, Tokyo 125-8585, Japan}

\author{Yudai Hirai}
\affiliation{Department of Applied Physics, Tokyo University of Science, Katsushika, Tokyo 125-8585, Japan}

\author{Rintaro Miyabayashi}
\affiliation{Department of Applied Physics, Tokyo University of Science, Katsushika, Tokyo 125-8585, Japan}

\author{Goro Shibata}
\affiliation{Department of Applied Physics, Tokyo University of Science, Katsushika, Tokyo 125-8585, Japan}
\affiliation{Materials Sciences Research Center, Japan Atomic Energy Agency, Sayo, Hyogo 679-5148, Japan}

\author{Hideki Kuwahara}
\affiliation{Department of Engineering and Applied Sciences, Sophia University, Chiyoda, Tokyo 102-8554, Japan}

\author{Miho Kitamura}
\affiliation{National Institutes for Quantum Science and Technology (QST), Sendai, Miyagi 980-8579, Japan}
\affiliation{Institute of Materials Structure Science, High Energy Accelerator Research Organization (KEK), Tsukuba, Ibaraki 305-0801, Japan}

\author{Koji Horiba}
\affiliation{National Institutes for Quantum Science and Technology (QST), Sendai, Miyagi 980-8579, Japan}
\affiliation{Institute of Materials Structure Science, High Energy Accelerator Research Organization (KEK), Tsukuba, Ibaraki 305-0801, Japan}

\author{Kenichi Ozawa}
\affiliation{Institute of Materials Structure Science, High Energy Accelerator Research Organization (KEK), Tsukuba, Ibaraki 305-0801, Japan}

\author{Noriaki Hamada}
\affiliation{Department of Physics and Astronomy, Tokyo University of Science, Noda, Chiba 278-8510, Japan}
\affiliation{R$^3$ Institute of Newly-Emerging Science Design, Osaka University, Toyonaka, Osaka 560-8531, Japan}

\author{Tomohiko Saitoh}\email[]{t-saitoh@rs.tus.ac.jp}
\affiliation{Department of Applied Physics, Tokyo University of Science, Katsushika, Tokyo 125-8585, Japan}

\date{\today}

\begin{abstract}

We have investigated the electronic structure of LaSr$_2$Mn$_2$O$_7$ in the CE-type antiferromagnetic (CE-AFM) state below the N\'eel temperature using micro-focused angle-resolved photoemission spectroscopy ($\mu$-ARPES) and tight-binding models. In agreement with the tight-binding calculations, we found a dispersive intensity around the X point in the $\mu$-ARPES spectra, where the A-type antiferromagnetic (A-AFM) phase has no bands experimentally and theoretically, demonstrating that we have successfully captured the signatures of the CE-AFM band structure for the first time. Many observed features can be explained by the CE-AFM tight-binding bands, although some of them and the overall near-Fermi level intensity mapping can be explained by the A-AFM band structure. This indicates that the CE-AFM domain size would be no larger than the beam footprint size of a $\sim\!20$~{\textmu}m scale.
\end{abstract}


\maketitle


\section{Introduction}
Transition-metal (TM) oxides have various intriguing properties that 
can be applied in the current and future smart society, for example, Ti oxides for photocatalysts \cite{fujishima1972electrochemical,FUJISHIMA2008515}, Co oxides for cathode materials of rechargeable batteries \cite{MIZUSHIMA1980783,goodenough2013li}, or high-temperature superconducting Cu oxides for zero-power loss transmission \cite{bednorz1986possible,RevModPhys.72.969}.
Many of such properties are manifestations of strong electron correlation between the TM $3d$ electrons and also complex interplay among the charge, spin, and orbital degrees of freedom. 

Giant or colossal magnetoresistance (GMR or CMR), a phenomenon of very large or huge reduction in resistivity by an applied magnetic field \cite{Baibich1988,Binasch1989,Chahara1993,vonHelmolt1993,tokura1994giant,moritomo1996giant}, 
is one of such properties with potential applications for spintronics devices or even a next-generation spacecraft radiator \cite{tachikawa2003development,tachikawa2022advanced}.
The CMR phenomenon is typically observed in the perovskite-type or the layered perovskite-type manganese oxides, more specifically, a certain $x$ range around $x\!=\!0.4$ of La$_{2-2x}$Sr$_{1+2x}$Mn$_2$O$_7$ (LSMO327), an $n=2$ member of the Ruddlesden-Popper (RP) type manganese oxides $A_{n+1}$Mn$_n$O$_{3n+1}$ \cite{moritomo1996giant}.

So far, the CMR phenomenon and its electronic structure have been the subject of many studies \cite{moritomo1996giant,Dessau1998,kimura2000layered,chuang2001fermi,Mannella2005,de2007quasiparticles}.
It is well-known that the CMR phenomenon basically originates from the double-exchange (DE) electronic structure \cite{doubleexchange}, namely the combination of a strong Hund's rule coupling between mobile charge carriers (holes/electrons in the Mn $3d$ $e_g$ state) and the large $S$=3/2 local spin of the Mn $3d$ $t_{2g}^3$ state that yields a ferromagnetic (FM) and metallic state.
However, because of the above complex interplay of multiple degrees of freedom, such as the superexchange interaction between $t_{2g}$ local spins, the Jahn-Teller instability, 
the orbital degrees of freedom, or charge ordering,
the situation in the real manganites is far more complicated.
As a result, a rich phase diagram emerges particularly near $x\!=\!0.5$ of LSMO327.
In this sense, a deep understanding of the electronic structure at and close to  $x\!=\!0.5$ directly contributes to understanding of the microscopic origin of the CMR phenomenon at $x \approx 0.4$, thereby systematically understanding the electronic structure of the manganese oxides.

\begin{figure}[h]
\centering
\includegraphics[clip,width = 7.5cm]{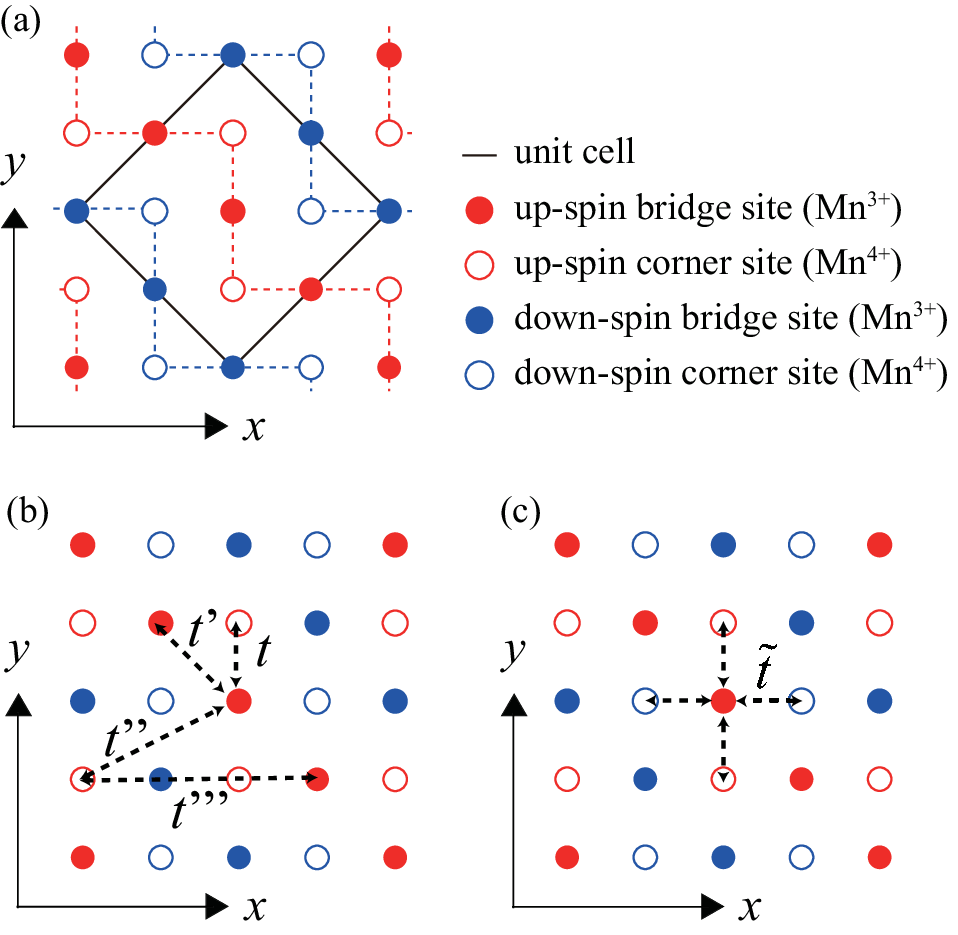}
\caption{(a) Schematic drawing of the CE-AFM spin-charge ordering. The up (down) spin sites are shown by the red (blue) circles and the Mn$^{3+}$ (Mn$^{4+}$) sites are shown by the filled (open) circles.
The zigzag FM chains are shown by red and blue dashed lines. 
(b) The hopping integrals in Model 1. Up to the 3rd nearest neighbors are considered. (c) The hopping integrals in Model 2. Only the nearest-neighbor FM and AFM couplings are considered.
For employing the same $\tilde{t}$ for both couplings, see Ref.~\onlinecite{Note1}.
}
	\label{zigzag}
\end{figure}

LSMO327 at $x\!=\!0.5$ is known to show the CE-type antiferromagnetic state (CE-AFM) that has a characteristic zigzag spin arrangement as shown in Fig.~\ref{zigzag}(a)
\cite{goodenough1955theory,kubota1999interplay,kubota1999neutron,Kubota2000RelationBC,dho2001re,rosciszewski2008electron}.
The important feature of CE-AFM is that
it is accompanied by the checkerboard-type charge ordering (CO) with a specific type of orbital ordering (OO)
\cite{kubota1999interplay,Sternlieb1996,Murakami1998}.
In LSMO327 at $x\!=\!0.5$, the charge-orbital ordered (COO) state appears below the COO transition temperature ($T_{\rm{COO}}) \sim\!200\!-\!210$ K and the CE-AFM state appears
below the CE-AFM transition temperature ($T_{\rm{N}}) \sim\!130\!-\!145$ K \cite{kubota1999interplay, Li2007}.
The CE-AFM phase of LSMO327 at $x\!=\!0.5$ appears only in a very narrow range of $x$ just around 0.50, which is surrounded by the A-type antiferromagnetic (A-AFM) phase with a tendency of phase separation into these two phases \cite{Li2007}.
Therefore, the CE-AFM phase easily collapses due to a small deviation of $x$ from 0.50 and/or a small amount of oxygen vacancies \cite{zheng2008}.
By contrast, the CE-AFM phase
is stable in La$_{0.5}$Sr$_{1.5}$MnO$_4$ (LSMO214 at $x\!=\!0.5$, an $n$\!=\!1 member of RP series) and was confirmed earlier than in LSMO327 at $x\!=\!0.5$ \cite{Sternlieb1996,Murakami1998}. However, precisely because of that, LSMO214 at $x\!=\!0.5$ is quite insulating at low temperatures and hence it is challenging to observe the CE-AFM band structure using angle-resolved photoemission spectroscopy (ARPES); the pioneering ARPES study on the CE-AFM systems was performed on LSMO214 at $x\!=\!0.5$ below $T_{\rm{COO}} =230$ K but above $T_{\rm{N}} \approx 110$ K of CE-AFM to minimize the charging effects \cite{Evtushinsky2010}. They studied the Fermi surface nesting properties in the COO state but did not deal much with the band dispersions.
Therefore, experimental observation of the electronic structure of the CE-AFM phase with COO has still been a long-standing problem .

On the theory side, many first-principles band-structure calculations have been performed for the FM and A-AFM states of LSMO327 \cite{de1999electronic,saniz2008orbital,sun2013minority}.
However,
there have been no first-principles band-structure calculations of the CE-AFM state in LSMO327 at $x\!=\!0.5$ so far because of the large unit cell of the CE-AFM state. 
Even on LSMO214 at $x\!=\!0.5$, there are only a few calculations with no report on the band dispersions \cite{Terakura1999,Solovyev2001}.
In the absence of realistic band-dispersion calculations, Solovyev and co-workers performed tight-binding calculations using effective Hamiltonians that
include only the $e_g$ orbitals \cite{Solovyev2001, Solovyev1999, Solovyev2003}.
They pointed out the importance of DE and OO in the zig-zag FM chain in the CE-AFM state \cite{Solovyev1999,Solovyev2001} and revealed the essential $e_g$ band dispersions \cite{Solovyev2003}.
More recently, tight-binding calculations for CO/OO and CE-AFM state for LSMO214 at $x\!=\!0.5$ were performed under several conditions \cite{Singh2016}, essentially reproducing the preceding CE-AFM band dispersions \cite{Solovyev2003}.
Nevertheless, validity of these calculations remains unknown because of the lack of experimental band dispersions.

In recent years, a significant progress has been made in ARPES, an experimental technique that can directly observe the electronic structure of materials. 
In particular, micro-focused ARPES ($\mu$-ARPES) is a very powerful tool because it enables us to perform ARPES measurements of very small samples and/or very small areas/domains of a variety of samples \cite{kitamura2022development,sugawara2023direct}. 

In this paper, we experimentally and theoretically explore the CE-AFM electronic structure of LSMO327 at $x\!=\!0.5$ using the $\mu$-ARPES technique and two tight-binding model calculations.

\section{Calculation and experiment}
\subsection{Calculation}

In the strong limit of Hund's rule coupling, the CE-AFM state can be reduced into the minimal model that only includes one FM zigzag chain of either up- or down-spins \cite{Solovyev1999,Terakura1999,Solovyev2001}. 
Based on this concept, we construct two tight-binding models (Model 1 and Model 2) for the CE-AFM state of LSMO327 at $x\!=\!0.5$.
Both are single-layered (namely, purely two-dimensional) models assuming the zigzag spin arrangement of CE-AFM.
We consider the two $e_g$ orbitals per Mn site
on the four inequivalent Mn sites in the unit cell of CE-AFM (see Fig.~\ref{zigzag}(a)), meaning 2 orbitals $\times$ 4 sites = 8 basis orbitals and thus we deal with $8\times8$ Hamiltonian matrices.  The four inequivalent sites can further be categorized into the two ``bridge sites" and the two ``corner sites" \cite{Solovyev2001}, which correspond to the Mn$^{3+}$ and Mn$^{4+}$ sites, respectively (Fig.~\ref{zigzag}(a)). We will use these terms hereafter.

Model 1 considers the hopping integrals of only the FM coupling but  including the long-range ones ($t', t''$ and $t'''$ in Fig.~\ref{zigzag}(b)).
This model is a natural extension of the model by Baublitz {\it et al.}
\cite{Baublitz2014}. They performed minimal but precise tight-binding band-structure calculations of the FM state of LSMO327 at $x\!=\!0.5$ and 0.38.
By switching off the hopping integrals between the up- and down-spin sites  in Fig.~\ref{zigzag}(a), we can extract Model 1 from their FM model.
Therefore, by adopting the same hopping integrals as Baublitz {\it et al.} used,
Model 1 can examine the phase transition from A-AFM to CE-AFM in the phase diagram at or close to $x\!=\!0.5$ \cite{Li2007}. 

To evaluate the effects of the AF coupling, we also constructed another tight-binding model (Model 2) that includes the FM and AF coupling but only considers the nearest neighbors (Fig.~\ref{zigzag}(c)). 
In Model 2, we also consider two $e_g$ orbitals represented by two tight binding parameters $\tilde{t}_{dd\sigma}$ (-0.652 eV) and $\tilde{t}_{dd\delta}$ (-0.011 eV), the values of which were determined from tight-binding fittings to the local-density approximation (LDA) calculations for $\rm{LaMnO_3}$ assuming the FM state. 
The exchange splitting energy was set to 3 eV \cite{Note1}.

For comparison with the A-AFM and FM electronic structure,
we also performed the band structure calculation on the A-AFM state \cite{kubota1999neutron,Kubota2000RelationBC} at $x\!=\!0.5$ and the FM state at $x\!=\!0.4$ using the code ABCAP with the full-potential linearized augmented plane-wave method in the LDA+$U$ scheme.
The effective on-site Coulomb repulsion $U_{\rm{eff}} = U - J$ was set to 2.0 eV for Mn $3d$ orbitals with $J=0.7$ eV \cite{Note4}.

\begin{figure}[t]
	\centering
	\includegraphics[clip,width = 8.5cm]{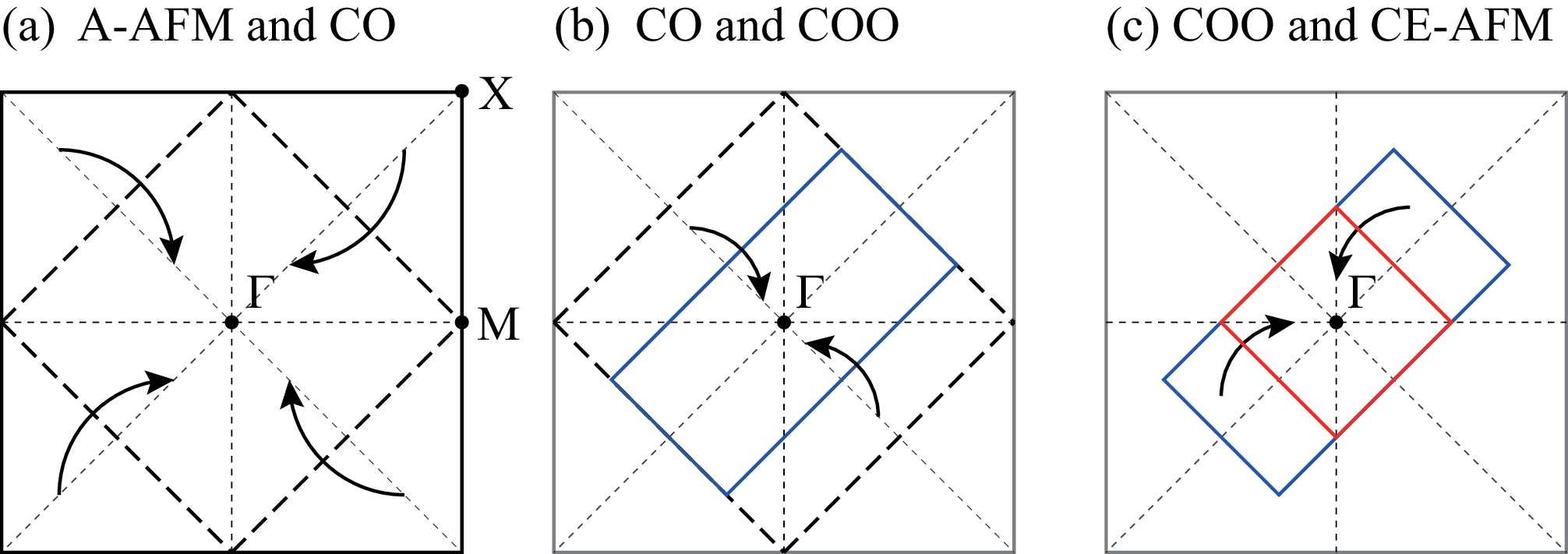}
	\caption{Comparisons of various BZs. Arrows show BZ-folding processes (a) The CO BZ (dashed square) compared with the A-AFM BZ (black square). (b) The COO BZ (blue rectangle) compared with the CO BZ. (c) The CE-AFM BZ (red square) compared with the COO BZ.}
	\label{BZFolding}
\end{figure}

Figure~\ref{BZFolding} shows how the Brillouin zone (BZ) of the A-AFM is folded into the CO, COO and CE-AFM BZ. In the CO and COO states, the X point is identical to the $\Gamma$ point, but the M point is still on the zone boundary in both BZs (Fig.~\ref{BZFolding}(a) and (b)). In the CE-AFM BZ, both the X and M points are identical to the $\Gamma$ point (Fig.~\ref{BZFolding}(c)). Note that although the CE-AFM BZ looks four-fold symmetric, it is actually two-fold symmetric because the four-fold symmetry is already broken in the COO BZ
due to the two-fold-symmetry nature of the OO and the FM spin ordering along the zigzag chains.

\begin{figure}[h]
	\centering
	\includegraphics[clip,width = 8cm]{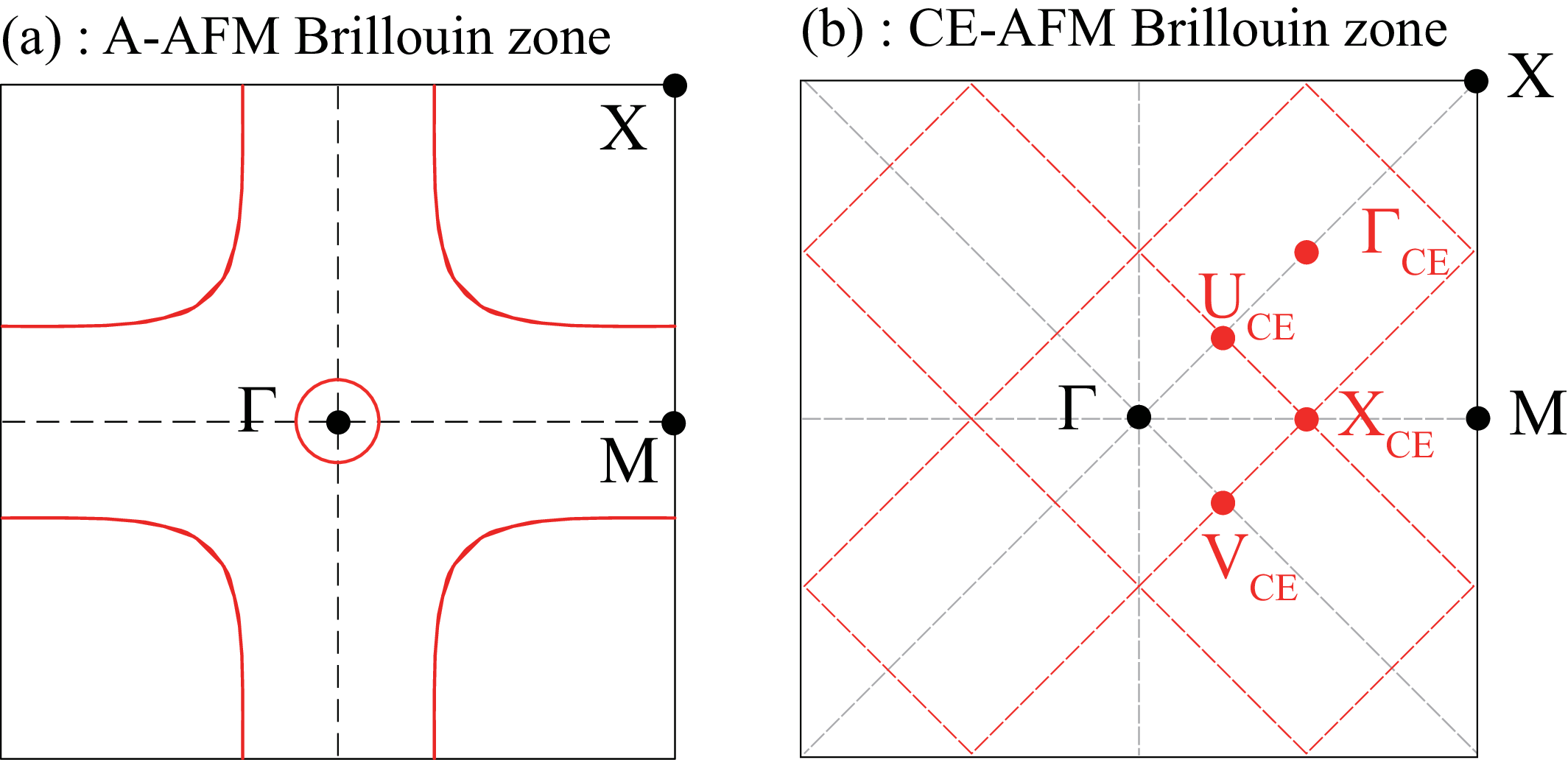}
	\caption{(a) Brillouin zone of the A-AFM state with the calculated Fermi surface of the A-AFM state at $x\!=\!0.5$. (b) Brillouin zones of the CE-AFM state (red) shown in Fig.~\ref{zigzag}(a) compared to that of A-AFM (black). For more details, see text.}
	\label{BZ}
\end{figure}

The A-AFM BZ with the calculated Fermi surface is shown in Fig.~\ref{BZ}(a).  In the CE-AFM state, this BZ is folded into the 1/8 area centered at the $\Gamma$ point (Fig.~\ref{BZ}(b) and also Fig.~\ref{BZFolding}(c)). As mentioned above, the M and X points in the A-AFM BZ are $\Gamma$ points in the CE-AFM BZ ($\rm{\Gamma_{CE}}$). The $k_x$ direction and $k_y$ direction are no longer equivalent in the CE-AFM BZ.
For example, $\rm{V_{CE}}$ and $\rm{U_{CE}}$ are not identical in the CE-AFM BZ. To distinguish the two lattices, we describe the symmetry points in the CE-AFM BZ with the subscript of ``CE" as shown in Fig.~\ref{BZ}(b).

\subsection{Experiment}
Single crystals of LSMO327 at $x\!=\!0.50$ were grown by the floating zone method.
To maximize the effect of oxygen annealing, the samples were crushed into thin flakes. The thin flake samples were annealed in oxygen atmosphere at 600$^\circ$C for 60 hours, and cooled at a rate of 0.5$\rm{{ }^\circ C/min}$.
This protocol of post-oxygen annealing followed by slow cooling has been reported to minimize oxygen vacancies and promote the growth of the CE-AFM phase \cite{zheng2008}.
In addition to LSMO327 $x\!=\!0.50$ samples, A-AFM $x\!=\!0.4$ samples were also grown using the floating zone method for comparison.
To indicate that our $x\!=\!0.5$ samples have been annealed, we hereafter describe them as ``0.50". The $x\!=\!0.4$ samples were not annealed because no precise control of oxygen content is required.
$\mu$-ARPES measurements were performed at a micro-focused beamline BL-28A in the Photon Factory, KEK by using a Scienta Omicron DA30 electron analyzer \cite{kitamura2022development}. 
Samples were cleaved {\it in situ} under the ultrahigh vacuum of $2 \times 10^{-8}$~Pa. The temperature was kept at 50 K well below $T_{\rm{N}}$ during the cleaving and the measurements.
At this temperature, only the A-AFM and CE-AFM phases exist, enabling us to identify the CE-AFM electronic structure by comparison with the A-AFM one.
The measurement cuts were along $\Gamma\!-\!\rm{X}$ and $\Gamma\!-\!\rm{M}$ for the $x\!=\!0.50$ samples and along $\Gamma\!-\!\rm{X}$ for the $x\!=\!0.4$ samples
with the photon energy of $h\nu\!=\!56$ eV, which is the Mn $3p\!-\!3d$ on-resonance excitation energy.
The beam spot size (horizontal $\times$ vertical) was 
 20~{\textmu}m~$\times$~20~{\textmu}m ($\Gamma\!-\!\rm{X}$ cut) and 
 10~{\textmu}m~$\times10$~{\textmu}m ($\Gamma\!-\!\rm{M}$ cut) for the $x\!=\!0.50$ sample, 
 and 
 28~{\textmu}m~$\times$~20~{\textmu}m ($\Gamma\!-\!\rm{X}$ cut) for the $x\!=\!0.4$ sample \cite{kitamura2022development}.

\section{Results and discussion}
\subsection{Tight-binding calculations}

\begin{figure*}[t]
	\centering
	\includegraphics[clip,width = 17cm]{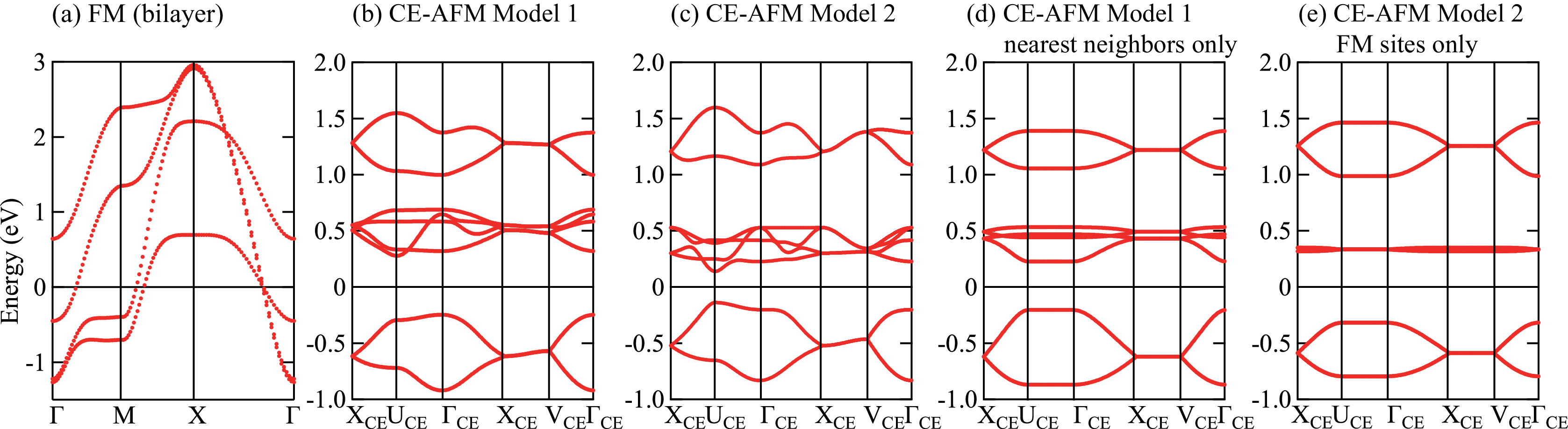}
	\caption{(a) Tight-binding majority-spin band structure of the bilayer FM state in Model 1 that reproduces the calculations in Ref.~\onlinecite{Baublitz2014}. (b) and (c) Tight-binding band structure of the CE-AFM state in Model 1 [(b)] and Model 2 [(c)]. (d) The same calculation as (b) including the nearest-neighbor hoppings only. (e) The same calculation as (c) considering the FM sites only.}
	\label{band}
\end{figure*}

First, we confirm that our Model 1 appropriately reduces into the model by Baublitz {\it et al.} \cite{Baublitz2014} by including the hopping integrals between up- and down-spin sites as well as the interlayer hopping. The result is shown in Fig. \ref{band}(a), which reproduces the Baublitz {\it et al.}'s result with the same hopping parameters \cite{Baublitz2014} and also agrees with many other preceding first-principles calculations \cite{de1999electronic,saniz2008orbital,PhysRevB.62.13318}. 
There are two issues to be noted:
(1) Because we consider only the single MnO$_2$ layer, the FM and A-AFM calculations are identical in our model.
(2) There is no near-Fermi level ($E_{\rm{F}}$) band around the X point. This will be a key to identify the experimental CE-AFM electronic structure later.

Next, we discuss the CE-AFM band structure.
Fig.~\ref{band}(b) shows the tight-binding band structure of the CE-AFM state in Model 1. 
The band dispersion in the CE-AFM state is quite different from that in the A-AFM state; the bands have a short dispersion period, and are flat and less-dispersive within $\sim\!0.7$~eV, reflecting the folding of the A-AFM BZ and the quasi one-dimensional nature of the zigzag spin chains.
The lower two bands below 0 eV are the bonding states while the upper two bands are their anti-bonding counterparts. Between the bonding and anti-bonding bands, there exist the four less-dispersive non-bonding bands.

Fig.~\ref{band}(c) shows the result of Model 2. 
One can see that Models 1 and Model 2 give similar band dispersions. 
This can be interpreted as that
the large-energy-scale band structure is primarily determined by the nearest-neighbor intra-chain coupling term, $t$ in Model 1 and the FM coupling of $\tilde{t}$ in Model 2.
The second-order hopping of $\tilde{t}$ at the corner sites of the chains in Model 2 gives the inter-chain interaction and corresponds to the inter-chain hopping terms $t''$ and $t'''$ in Model 1. 
These long-range interactions describe weak modulations of the band structure, determining the band dispersions in the smaller energy scale.
Relatively small differences in Figs.~\ref{band}(b) and \ref{band}(c) originate from small differences of the hopping integrals in size and character.
Both in Model 1 and Model 2, the bonding and anti-bonding bands on
the $\Gamma_{\rm{CE}}\!-\!\rm{U}_{\rm{CE}}$ line
are less dispersive than those on
the $\Gamma_{\rm{CE}}\!-\!\rm{V}_{\rm{CE}}$ line. 
This is because the direction of the FM zigzag chains, $(0,0)-(\pi,-\pi)$, is the same as
the $\Gamma_{\rm{CE}}\!-\!\rm{V}_{\rm{CE}}$ line
and perpendicular to 
the $\Gamma_{\rm{CE}}\!-\!\rm{U}_{\rm{CE}}$ line
(see Figs.~\ref{zigzag} and \ref{BZ}). 
This interpretation also holds for the bands on
the $\rm{X}_{\rm{CE}}\!-\!\rm{U}_{\rm{CE}}$ line.

All the above results indicate that the essence of the CE-AFM band structure is the quasi one-dimensional FM chain as pointed out by Solovyev and coworkers \cite{Solovyev1999,Terakura1999,Solovyev2001}.
This can be confirmed by considering
only the intra-chain FM hopping term in both models. 
The results are shown in Figs.~\ref{band}(d) and \ref{band}(e).
In this limit, Model 1 and Model 2 are identical.
We notice that all the bands on
the $\Gamma_{\rm{CE}}\!-\!\rm{U}_{\rm{CE}}$ line
and
the $\rm{X}_{\rm{CE}}\!-\!\rm{U}_{\rm{CE}}$ line
are completely flat in Figs.~\ref{band}(d) and \ref{band}(e),
due to no hopping term along the $(\pi,\pi)$ direction.
The non-bonding bands are essentially non-dispersive and should be completely flat in the limit of $dd\delta \rightarrow 0$ at the bridge sites, which describes the complete $(3x^2-r^2/3y^2-r^2)$-type OO \cite{Solovyev1999}. 
These features agree well with the preceding works \cite{Solovyev2003, Singh2016}.
The minor differences between Figs.~\ref{band}(d) and \ref{band}(e), especially in the non-bonding bands, are due to the larger parameter values in Model 1.

\subsection{$\mu$-ARPES}

\begin{figure*}[t]
	\centering
	\includegraphics[clip,width = 16cm]{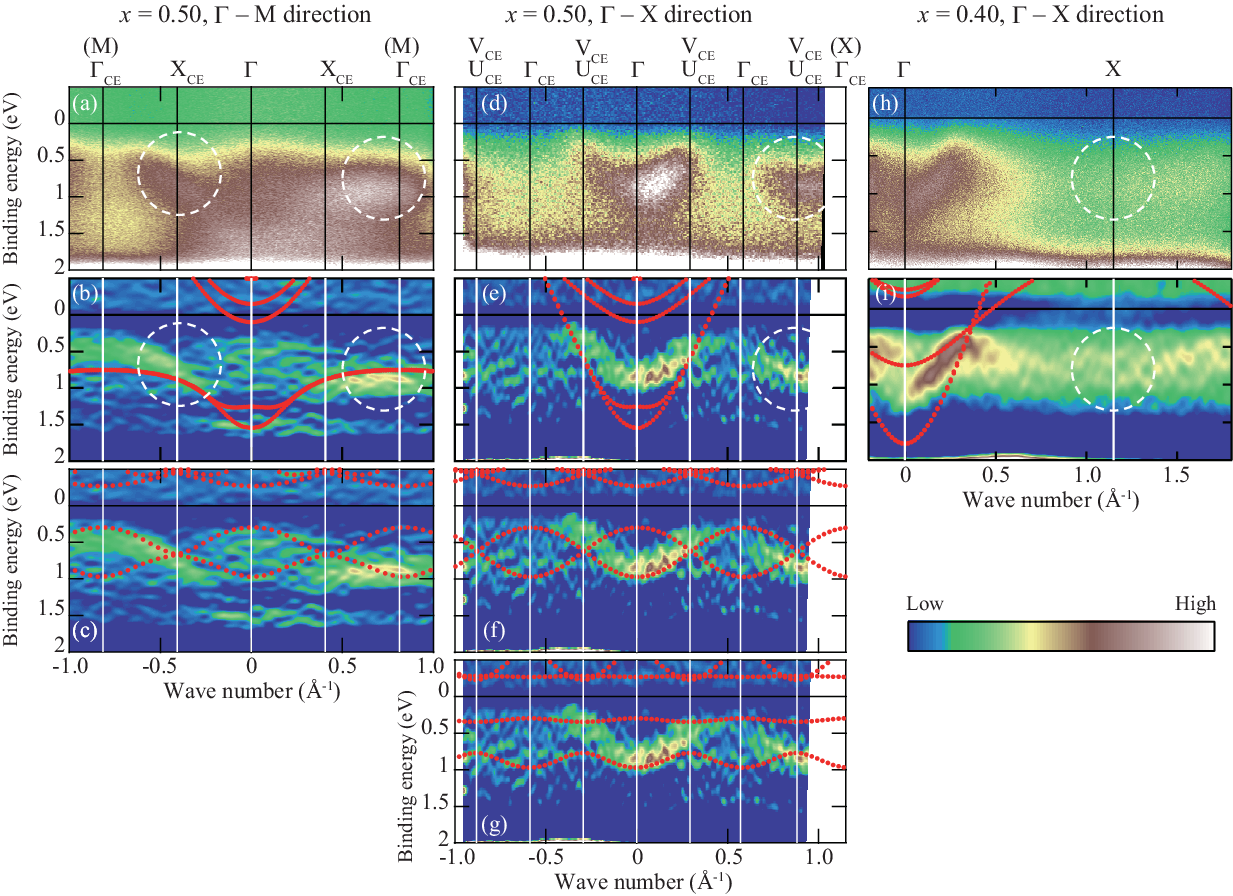}
	\caption{
		(a) The $\mu$-ARPES intensity plot of the $x\!=\!0.50$ sample along the $\Gamma\!-\!\rm{X}_{\rm{CE}}\!-\!\Gamma_{\rm{CE}}$ ($\Gamma\!-\!\rm{M}$) direction.  White dashed circles show significant intensity around the right $\Gamma_{\rm{CE}}$ (M) and the left X$_{\rm{CE}}$ points.
		(b) A second-derivative plot of Panel (a) compared with the band-structure calculation of the A-AFM state at $x\!=\!0.50$ (red dotted lines).
		(c) The second-derivative plot compared with the tight-binding calculation in Model 1 (red dotted lines).
		(d) The $\mu$-ARPES intensity plot of the $x\!=\!0.50$ sample along the $\Gamma\!-\!\rm{V/U}_{\rm{CE}}\!-\!\Gamma_{\rm{CE}}$ ($\Gamma\!-\!\rm{X}$) direction. A white dashed circle shows a hot spot around the $\Gamma_{\rm{CE}} (\rm{X})$ point. 
		(e) A second-derivative plot of Panel (d) compared with the band-structure calculation of the A-AFM state at $x\!=\!0.50$ (red dotted lines).
		(f) The second-derivative plot compared with the tight-binding calculation along the $\Gamma_{\rm{CE}}\!-\!\rm{V}_{\rm{CE}}$ line in Model 1 (red dotted lines).
		(g) The second-derivative plot compared with the tight-binding calculation along the $\Gamma_{\rm{CE}}\!-\!\rm{U}_{\rm{CE}}$ line in Model 1 (red dotted lines).
		(h) The $\mu$-ARPES intensity plot of the $x\!=\!0.4$ sample along the $\Gamma\!-\!\rm{X}$ direction. A white dashed circle highlights no intensity around the X point. 
		(i) A second-derivative plot of Panel (h), compared with the band-structure calculation of the FM state at $x\!=\!0.4$ (red dotted lines).
		In the band-structure calculations in Panels (b), (e), and (i), only the $e_g$ bands are shown for clarity.
	}
	\label{result}
\end{figure*}

Figure~\ref{result} shows the result of our $\mu$-ARPES measurements of the $x\!=\!0.5$ sample along the two symmetry lines together with the result of the $x\!=\!0.4$ sample along the $\Gamma\!-\!\rm{X}$ line.
Again, please note that
the $\Gamma\!-\!\rm{M}$ and the $\Gamma\!-\!\rm{X}$ lines in the A-AFM Brillouin zone correspond to the
$\Gamma(\Gamma_{\rm{CE}})\!-\!\rm{X}_{\rm{CE}}\!-\!\Gamma_{\rm{CE}}$ and
$\Gamma(\Gamma_{\rm{CE}})\!-\!\rm{U}_{\rm{CE}}\!-\!\Gamma_{\rm{CE}}\!-\!\rm{U}_{\rm{CE}}\!-\!\Gamma_{\rm{CE}}$ lines in the 
CE-AFM Brillouin zone, respectively (see also Fig.~\ref{BZ}).

Figure~\ref{result}(a) shows the $\mu$-ARPES intensity plot along the  $\Gamma\!-\!\rm{X}_{\rm{CE}}\!-\!\Gamma_{\rm{CE}}$ ($\Gamma\!-\!\rm{M}$) direction.
A shallow parabolic-like band with the vertex at $\sim\!1.5$ eV at the $\Gamma$ point can be observed. In addition, one can observe significant spectral intensity around the two $\Gamma_{\rm{CE}} (\rm{M})$ points on the both sides of $\Gamma$ (highlighted by white dashed circles). In particular, a hot spot around the right $\Gamma_{\rm{CE}} (\rm{M})$ ($k\sim\!0.8$~\AA$^{-1}$) is remarkable. 
Note that the non-dispersive strong intensity below $\sim\!1.7\!-\!1.8$ eV (also in Figs.~\ref{result}(d) and (h)) is due to the $t_{2g}$ bands.

To see the experimental band dispersion in more detail, a second-derivative plot of Fig.~\ref{result}(a) is shown in Fig.~\ref{result}(b) together with the theoretical $e_g$ bands calculated with LDA+$U$ band-structure calculation of the A-AFM state at $x\!=\!0.50$ (the red dotted lines). Now the experimental band structure is more clear. The dispersion is found to be rather narrow within $\sim\!0.7$~eV between the binding energy of $\sim\!0.4$~eV and $\sim\!1.1$~eV. 
The band around the right $\Gamma_{\rm{CE}}$ (M) point looks rather non-dispersive (the right white dashed circle). 
Also, one can clearly see the band from the left $\Gamma_{\rm{CE}}$ (M) point through the left X$_{\rm{CE}}$ point, probably continuing to the center $\Gamma$ point.

The A-AFM band structure may explain the overall experimental band dispersion, particularly the non-dispersive band around the right $\Gamma_{\rm{CE}}$ (M) point and the intensity at $\sim\!1.5$~eV around the $\Gamma$ point.
However, it deviates from the band from the left $\Gamma_{\rm{CE}}$ through the left X$_{\rm{CE}}$ point.
The symmetric A-AFM band cannot explain the bands around the two $\Gamma_{\rm{CE}}$s
because the binding energies of the two bands are different ($\sim\!1$~eV around the right $\Gamma_{\rm{CE}}$ point and $\sim\!0.5$~eV around the left $\Gamma_{\rm{CE}}$ point.) 
On the other hand, the CE-AFM tight-binding band structure of Model 1 shown in Fig.~\ref{result}(c) (the red dotted lines) reproduces these features; the experimental band width of $\sim\!0.5$~eV matches the calculation and the bands around both the right and left $\Gamma_{\rm{CE}}$ points can be explained by a part of the CE-AFM band structure.
In addition, the broad intensity around the $\Gamma$ point between $\sim\!0.5\!-\!1.0$ eV can be explained by the CE-AFM bands.

Figure~\ref{result} (d) shows the $\mu$-ARPES intensity plot along the  $\Gamma\!-\!\rm{V/U}_{\rm{CE}}\!-\!\Gamma_{\rm{CE}}$ ($\Gamma\!-\!\rm{X}$) direction.
In addition to a parabolic band with the vertex at about 1 eV centered at the $\Gamma$ point, there can be seen a dispersive intensity from the $\rm{V/U}_{\rm{CE}}$ point to the $\Gamma_{\rm{CE}} (\rm{X})$ point near $k=1.0$~\rm{\AA}$^{-1}$.
In Fig.~\ref{result}(e), a second-derivative plot of Fig.~\ref{result}(d) is shown together with the theoretical $e_g$ bands calculated with LDA+$U$ band-structure calculation of the A-AFM state at $x\!=\!0.50$ (the red dotted lines). 
Although the A-AFM band structure can explain the parabolic band at the $\Gamma$ point, its vertex location is rather too low. 
Although this disagreement may be explained by a band renormalization due to the strong electron-electron/electron-phonon coupling,
it can never reproduce the band near $k=1.0$~\rm{\AA}$^{-1}$ because there is no $e_g$-band around the X point. We thus compare the experiment with the CE-AFM tight-binding calculations of Model 1.
Because we cannot experimentally distinguish the $\Gamma\!-\!\rm{V}_{\rm{CE}}$ and the $\Gamma\!-\!\rm{U}_{\rm{CE}}$ directions, we compare the experiment with each direction below, respectively.

Figure~\ref{result}(f) compares the second-derivative plot with the CE-AFM tight-binding band structure (the red dotted lines) along the $\Gamma\!-\!\rm{V}_{\rm{CE}}$ direction.
The parabolic band at the $\Gamma$ point is reproduced well and the band near $k=1.0$~\rm{\AA}$^{-1}$ can be explained as a part of the band from the second $\Gamma_{\rm{CE}}$ point ($\sim\!0.6$~\rm{\AA}$^{-1}$) down to the third $\Gamma_{\rm{CE}}$ point ($\sim\!1.2$~\rm{\AA}$^{-1}$: outside the experimental area)  through the $\rm{V}_{\rm{CE}}$ point.
On the other hand, the CE-AFM tight-binding band structure along the $\Gamma\!-\!\rm{U}_{\rm{CE}}$ direction in Fig.~\ref{result}(g) (the red dotted lines) does not explain as well as the band structure along the $\Gamma\!-\!\rm{V}_{\rm{CE}}$ direction.
Therefore, we believe that we have observed the band structure along the $\Gamma\!-\!\rm{V}_{\rm{CE}}$ direction in this experimental geometry.

To compare the above results with the experimental FM band structure, we also show the $\mu$-ARPES intensity plot of the $x\!=\!0.4$ sample along the $\Gamma\!-\!\rm{X}$ direction in Fig.~\ref{result}(h). It is apparent that no significant intensity can be observed around the X point as shown by the white dashed circle. 
A second-derivative plot of Fig.~\ref{result}(h) is shown in Fig.~\ref{result}(i) together with the LDA+$U$ $e_g$ band of the FM state at $x\!=\!0.4$ (the red dotted lines). Like Fig.~\ref{result}(e), there is no $e_g$ band around the X point, which completely agrees with the experimental result.
We note that the observed bending-back behavior of the experimental band dispersion around $k\sim0.4$~\rm{\AA}$^{-1}$ in Fig.~\ref{result}(i) can be interpreted as the pseudogap, which is typically observed in the hole Fermi surface around the X point of the $x=0.4$ samples \cite{Dessau1998,Mannella2005}.

From the above considerations, 
we believe that we have successfully captured 
the signatures
of the experimental CE-AFM band structure for the first time although the A-AFM domain would be coexisting with it in the probed area.
Here we note that the ARPES study on LSMO214 at $x\!=\!0.5$ in the COO state (but \textit{not} in the CE-AFM state) did not observe any signature of a band around the X ($\Gamma_{\rm{CE}}$) point \cite{Evtushinsky2010},
where a replica band at the original $\Gamma$ point should appear even in the COO \cite{Singh2016} (see also Fig.~\ref{BZFolding}).
This discrepancy may be explained by the difference between the spin-disordered COO (above $T_{\rm{N}}$) and the spin-ordered CE-AFM state, as pointed out by Solovyev, although there still exists a broad band around the X ($\Gamma_{\rm{CE}}$) point in the calculations \cite{Solovyev2003}.
We also note that while the X ($\Gamma_{\rm{CE}}$) point is identical in both the COO and CE-AFM BZs, the M ($\Gamma_{\rm{CE}}$) point does not correspond to the $\Gamma$ point in the COO BZ; hence, they represent different $k$-space points. Although only a single band at the M ($\Gamma_{\rm{CE}}$) point in the COO calculation\cite{Singh2016}, we observed a lower- and a higher-binding-energy branches in the CE-AFM band structure around the right and left M ($\Gamma_{\rm{CE}}$) points, respectively (Fig.~\ref{result}(c)). This would further support the validity of the CE-AFM band structure.

\begin{figure}[b]
	\centering
	\includegraphics[clip,width = 8.5cm]{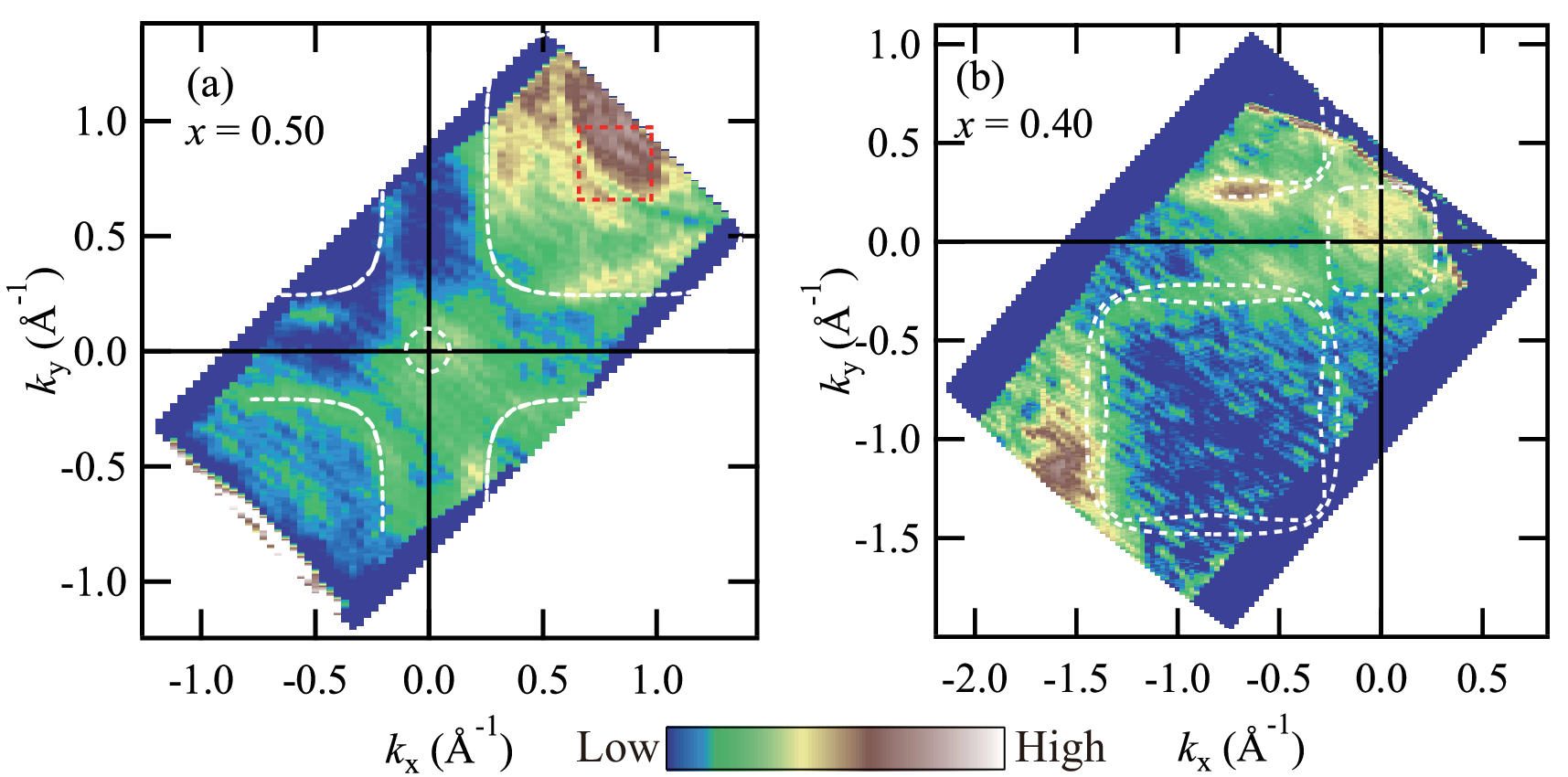}
	\caption{(a) Near-$E_{\rm{F}}$ intensity mapping of the $x\!=\!0.50$ sample compared with the calculated A-AFM  Fermi surface (white dashed lines). The red dashed square shows the intensity around the X ($\Gamma_{\rm{CE}}$) point that cannot be explained by the A-AFM Fermi surface.
	(b) Near-$E_{\rm{F}}$ intensity mapping of $x\!=\!0.4$ sample compared with the calculated FM Fermi surface (white dashed lines). The intensity in both panels is taken at $0.4\pm0.2$ eV of the binding energy.}
	\label{FSmap}
\end{figure}

Figure~\ref{FSmap}(a) compares the near-$E_{\rm{F}}$ intensity of the $x\!=\!0.50$ sample with the calculated A-AFM Fermi surface (white dashed lines) using LDA+$U$ method. Because there is no appreciable intensity at $E_{\rm{F}}$ in experiment, we show the intensity mapping taken at $0.4\pm0.2$ eV of the binding energy. 
The A-AFM Fermi surface explains the overall experimental intensity mapping,
although the data are partially blurred by intensity suppression or modulation, probably arising from the experimental geometry.
At the same time, however, an appreciable intensity can apparently be observed around the X ($\Gamma_{\rm{CE}}$) point at $(k_x,k_y)\approx(0.85,0.85)$ (highlighted by the red dashed square), which cannot be reproduced by the A-AFM band structure.
By contrast, the near-$E_{\rm{F}}$ intensity of the $x\!=\!0.4$ sample shown in Fig.~\ref{FSmap} is reasonably reproduced by the FM Fermi surface, including no appreciable intensity around the X point at $(k_x,k_y)\approx(-0.85,-0.85)$.
Therefore, we have probably observed both the CE-AFM and A-AFM domains in Figs.~\ref{result} and \ref{FSmap}(a) in the $x\!=\!0.50$ case. Because the footprint size of the beam is no larger than $\sim\!30$~{\textmu}m~$\times$~20~{\textmu}m \cite{Note2}, the typical CE-AFM domain would be smaller than this size, surrounded by the A-AFM domains. 
This estimation is consistent with a reported COO domain of LSMO214 at $x\!=\!0.5$ on the {\textmu}m scale \cite{Murakami2010,Note3} and also a recent study on the CE-AFM state of a LSMO327 $x\!=\!0.51$ sample with no post-oxygen annealing by using a scanning-type resonant soft x-ray scattering \cite{Nakao2025,Note5}.

\section{Conclusions}

We have investigated the CE-AFM electronic structure of LSMO327 at $x\!=\!0.50$ using $\mu$-ARPES and tight-binding calculations. 
We constructed two tight-binding models for the CE-AFM state, which confirmed the importance of the DE FM zigzag chains with orbital ordering.
In the $\mu$-ARPES spectra of LSMO327 at $x\!=\!0.50$, we observed a distinct dispersive band around the X ($\Gamma_{\rm{CE}}$) point, where the A-AFM band structure has no bands as confirmed experimentally and theoretically in LSMO327 at $x\!=\!0.4$. By contrast, the CE-AFM tight-binding models predict narrow bands around the X ($\Gamma_{\rm{CE}}$) point, demonstrating that we have successfully captured the signatures of the experimental CE-AFM band structure.
At the same time, some of the band features and the overall near-$E_{\rm{F}}$ intensity mapping can be explained by the A-AFM calculations, indicating that the domain size of the observed CE-AFM state would be no larger than the beam footprint size of a $\sim\!20$~{\textmu}m scale.

\begin{acknowledgments}
The authors thank S. Souma and T. Sato for their experimental support and D. Ootsuki and T. Mizokawa for fruitful discussions.
This work was supported by JST, the establishment of university fellowships
towards the creation of science technology innovation, Grant Number JPMJFS2144, and by 
Japan Society for the Promotion of Science (JSPS) 
Grants-in-Aid for Scientific Research (KAKENHI) 
Nos. JP18K03543, JP18K03550, JP21H01536, JP21K03474, JP23K03326, and JP24K06963.
The synchrotron radiation experiments were performed under the approval of the Photon Factory Program Advisory Committee
(Proposal Nos. 2019G525, 2021S2-001, 2021G552, 2023G608, 2023G683, 2024S2-001).
\end{acknowledgments}

%

\end{document}